\documentclass[12pt]{article}
\usepackage{epsfig,amssymb}

\newcommand{\bq}{\begin{equation}}
\newcommand{\eq}{\end{equation}}
\newcommand{\bqa}{\begin{eqnarray}}
\newcommand{\eqa}{\end{eqnarray}}
\newcommand{\ben}{\begin{enumerate}}
\newcommand{\een}{\end{enumerate}}
\newcommand{\bc}{\begin{center}}
\newcommand{\ec}{\end{center}}
\newcommand{\bqb}{\begin{eqnarray*}}
\newcommand{\eqb}{\end{eqnarray*}}

\def\lsim{\lesssim}

\def\ie{{\it i.e. }}
\def\eg{{\it e.g. }}

\def\etal{{\it et.al. }}

\def\s2beta{s_{2 \beta}}
\def\c2beta{c_{2 \beta}}

\def\pr#1#2#3{ Phys. Rev. ${\bf{#1}}$: #2 (#3)}
\def\prl#1#2#3{ Phys. Rev. Lett. ${\bf{#1}}$: #2 (#3)}
\def\pl#1#2#3{ Phys. Lett. ${\bf{#1}}$: #2 (#3)}

\def\sjnp#1#2#3{ Sov. J. Nucl. Phys. ${\bf{#1}}$: #2 (#3)}
\def\mpl#1#2#3{ Mod. Phys. Lett. ${\bf{#1}}$: #2 (#3)}

\def\JHEP#1#2#3{JHEP ${\bf{#1}}$: #2 (#3)}

\def\APPol#1#2#3{ Acta Phys. Pol. ${\bf{#1}}$: #2 (#3)}
\def\Astropart#1#2#3{ Astropart. Phys. ${\bf{#1}}$: #2 (#3)}

\begin{document}
\pagenumbering{arabic} \thispagestyle{empty}
\def\thefootnote{\fnsymbol{footnote}}
\setcounter{footnote}{1}

\begin{flushright}
THES-TP 2002/03 \\
PM/02-55 \\
hep-ph/0212110 \\
Revised version,\\
 November  2003 \\
 \end{flushright}
\vspace{2cm}
\begin{center}
{\Large\bf  The flavor distribution of
Cosmic Neutrinos\footnote{Partially
supported by EU contract   HPRN-CT-2000-00149, and the
PLATON French-Greek Collaboration project, 2002.}.}
 \vspace{1.5cm}  \\
{\large G.J. Gounaris$^a$ and G.  Moultaka$^b$ }\\ \vspace{0.4cm}
$^a$Department of Theoretical Physics, Aristotle University of
Thessaloniki,\\ Gr-541 24, Thessaloniki, Greece.\\
\vspace{0.2cm}
$^b$Physique
Math\'{e}matique et Th\'{e}orique,
UMR 5825\\
Universit\'{e} Montpellier II,
 F-34095 Montpellier, France.\\

\vspace{2.cm}

{\bf Abstract }
\end{center}

\noindent
Simple approximate expressions for the relative flavor
fluxes  of energetic cosmic neutrinos detected on Earth,  are presented
 in terms of their initial fluxes at the surface of the producing  cosmic
 sites, assuming the neutrino mixing angles to lie  in the experimentally
favored region. These expressions highlight the main sensitivities
 to the initial production fluxes as well as to small variations of the
 mixing angles within the experimentally preferred region,
 thus providing simple methods to disentangle these
physical quantities from cosmic neutrino data. This is more so, due
to some striking mathematical  properties of the relative neutrino flavors
 at $(s_{23}=1/\sqrt{2}, ~s_{13}=0)$, which somehow
 characterize  the whole experimentally
 preferred region.
 We also assess the quality of our approximate expressions
through a numerical comparison with the exact results.

\vspace{1cm}
\noindent
PACS number(s): 14.60.Lm, 95.85.Ry, 96.40.Tv, 98.70.Sa. \\

\def\thefootnote{\arabic{footnote}}
\setcounter{footnote}{0}
\clearpage

\noindent
The relative flavors of energetic  Neutrinos reaching the Earth,
after they have been emitted at various cosmic sites, provide
useful information on the physical conditions there.
Such very energetic neutrinos, approaching   the $10^3$TeV
\cite{Waxman}, or even the $10^6$TeV scale
\cite{Halzen, Athar, Yasuda, Magnetar}, may  be generated in
 extra-galactic  sites like Gamma Ray Bursts and Active Galactic Nuclei
(AGN). Galactic candidates that may emit neutrinos of up to 100 TeV
have also been identified at distances of at
least $2.6 {\rm kpc}$   \cite{galactic-neutrinos, Magnetar}.
[Exploding galactic Supernovae may
also induce observable neutrino fluxes with energies
in the few MeV range \cite{Raffelt1}.]

It is commonly believed that these neutrinos are produced mainly
through the
decay of   high energy $\pi^\pm$ and K  (and may be also some D) mesons, which
implies that the initial relative neutrino flavors
 at the cosmic sites satisfy
\bq
F^0_e=1/3 ~, ~ F^0_\mu=2/3~, ~F^0_\tau=0~ ,
\label{canonical-initial-flavors}
\eq
\cite{Athar, Yasuda, Magnetar, Halzen, canonical-F0}.
This case is referred to below, as the
{\sl canonical case}. It may be useful to remember
though, that our present understanding of the mechanisms for generating
high energy neutrinos is rather primitive, and sites may exist in
the Universe where the produced neutrinos have
a different initial structure \cite{Halzen, King}.
 Therefore, the measurement of the relative intensities
 of the various neutrino flavors on Earth,
 may provide useful direct information on the mechanism
 responsible for their  generation in the Cosmos and may possibly
 lead to the discovery of some kind of New  Physics \cite{Shahid}.

Once  the various neutrino flavors produced inside some cosmic object
reach  its  surface,   they  propagate oscillating through space, as dictated
by the vacuum oscillation formalism\footnote{For a review see \eg
\cite{Neutrino-factory}.}. Some of the intrinsic properties of the flavor
oscillations are, however, not very easy to readout from the general
vacuum oscillation formulae,
especially when relying only on numerical scans. These properties can be
of importance in the perspective of disentangling, through the
measurement of relative fluxes on earth, the astrophysical
uncertainties encoded in the initial flavor fluxes on the cosmic sites,
from the particle physics features in the neutrino sector.
It will therefore be useful to have
simple formulae giving  the observable relative
neutrino flavors  $F_e, ~F_\mu, ~F_\tau$ on Earth,
in terms of the initially  produced
ones $F^0_e, ~F^0_\mu, ~F^0_\tau$ at the surface of the
cosmic object.\\

The aim of the present paper is to give such simple analytical expressions,
assuming only three active neutrino flavors which propagate oscillating
among themselves \cite{Neutrino-factory}; (in particular no cosmic neutrino
decay is assumed \cite{Beacom}). These expressions have the advantage of
identifying specific sensitivities and degeneracies in the mixing angles, and
of encoding some essential qualitative and quantitative properties of the
full formalism. Some of these properties
turn out to be highly non-generic consequences of the  concomitance
of the physically favored values for the mixing angles $s_{23}, s_{13}$, and
the initial fluxes.

To derive  the aforementioned formulae,
we  take into account the basic experimental
 characteristics of the neutrino masses and mixings.
These are summarized as follows: The recent SNO \cite{SNO}  data
combined with those of Super-Kamiokande \cite{SuperKamiokande} strongly
favor the LMA MSW   \cite{MSW} solar solution with three active
neutrinos and $\theta_{12}\simeq \pi/5.1$ and  $|m_2^2-m_1^2|\simeq
5 \times 10^{-5} \, eV^2$ \cite{solar}. The atmospheric
neutrino \cite{atmospheric-SuperKamiokande} data imply
$\theta_{23}\simeq \pi/4$ and $|m_3^2-m_2^2|\simeq
2.5 \times 10^{-3} \, eV^2$; while the CHOOZ experiment constrains
$\theta_{13} \lsim 0.1$, \cite{CHOOZ, Neutrino-factory}.
Defining $s_{ij} \equiv \sin \theta_{ij}$ and choosing
 the ``central values"
\bq
s_{12}^c = \frac{1}{\sqrt{3}} ~~~,~~~ s_{23}^c = \frac{1}{\sqrt{2}} ~~~,~~~
s_{13}^c =0 \label{central-values}
\eq
one has
\bq
s_{12} \equiv s_{12}^c +\delta s_{12} ~~~,~~~
s_{23} \equiv s_{23}^c +\delta s_{23}~~~,~~~
s_{13} \equiv s_{13}^c + \delta s_{13} \label{angle-range-def}
\eq

where \cite{solar,atmospheric-SuperKamiokande,Neutrino-factory, s12-limits},
\bqa
& -0.11 \lsim \delta s_{12} \lsim 0.04 &  ~~, ~~
-0.12 \lsim \delta s_{23} \lsim 0.10 ~~, \label{s12-s23-range} \\
&  0 \leq  \delta s_{13}  \lsim 0.1 & ~~, \label{s13-range}
\eqa

For  realistic  neutrino mass differences, and
  neutrino energies in the range
 $  E \lsim 10^6 \, {\rm TeV}$,
the vacuum  oscillation lengths $\lambda_{ij}= 4 \pi
E/|m_i^2-m_j^2|$,   always satisfy
$\lambda_{ij} \lsim 1\, {\rm pc}$,
which is much smaller than
the distances to all cosmic neutrino emitting
sites, beyond our solar system \cite{galactic-neutrinos}.
Consequently, the number of
oscillations performed  by   the cosmic
 neutrinos before arriving at the
Earth, is so  large, that
$\sin^2 (\pi L/ \lambda_{ij})$ averages  to $ 1/2$,  and
the CP-violating contributions vanish.

Studying the  properties of the relative neutrino fluxes
in  the $s_{23}s_{13}$-plane,
for any fixed values of $s_{12}$ and $\cos\delta$,
 we remark that the point $(s_{23}^c,s_{13}^c)=(1/\sqrt{2}, 0)$, (lying  within the experimental range of
  Eqs.(\ref{angle-range-def}, \ref{s12-s23-range}, \ref{s13-range})),
presents   the  striking property of being the  \underline{unique  point}  of
this $s_{23}s_{13}$-plane\footnote{At least for $s_{23}<0.5$.},  where
 a common ($s_{12}$- and $\cos\delta$-dependent) direction  exists,
 along which all
three fluxes $F_e,~ F_\mu, ~F_\tau$ are stationary.

As an example, we expand the standard vacuum oscillation formulae   to first
order in $s_{13}$  and $(\delta s_{12}, ~\delta s_{23})$ defined in
(\ref{angle-range-def}), getting the relative neutrino fluxes   on Earth,
\bqa
 F_e &=& \frac{1}{3} + \frac{(1 - 2 \sqrt{3} \delta s_{12})}{9} ( 3 F^0_e -1)
 +\delta s_{123} (F^0_\tau -F^0_\mu)  ~~ , \nonumber  \\
 F_\mu &=& \frac{1}{3} + \frac{( 2 \sqrt{3} \delta s_{12} -1 )}{18}
( 3 F^0_e -1)
 +\delta s_{123}   (F^0_\mu -F^0_e)  ~~ , \nonumber \\
 F_\tau &=& \frac{1}{3} + \frac{( 2 \sqrt{3} \delta s_{12} -1 )}{18}
( 3 F^0_e -1)
 +\delta s_{123}   (F^0_e -F^0_\tau)  ~~ , \label{relative-flavors}
\eqa
where
\bq
\delta s_{123} \equiv \frac{\sqrt{2}}{9}
 (\kappa \delta s_{23}-\delta s_{13} \cos\delta ) ~~, \label{a123-def}
\eq
with $\kappa = 4$,
and $(F^0_e,~  F^0_\mu, ~ F^0_\tau)$ being
the initial neutrino relative flavors
 at the cosmic site. The
experimental  constraints (\ref{s12-s23-range}, \ref{s13-range}) then imply
\bq
-0.09 \lsim \delta s_{123} \lsim +0.07 ~~ .\label{a123-range}
\eq
In writing Eq.(\ref{relative-flavors}) we took into account
the unitarity relation
\bq
F_e +F_\mu + F_\tau = F^0_e +F^0_\mu + F^0_\tau =1 ~~,
\label{normalization}
\eq
where   the right hand side  is just a normalization.

The set of equations (\ref{relative-flavors}) clearly indicates that
for $s_{12}^c\equiv 1/\sqrt{3}$, all three relative fluxes on Earth
are stationary along the direction
$ \kappa [s_{23}-1/\sqrt{2}]-s_{13} \cos\delta =0$ passing through
the point $(s_{23},s_{13})=(1/\sqrt{2}, 0)$ of
the $s_{23}s_{13}$-plane\footnote{For a given central value
$s_{12}^c$, the stationary direction is  determined by
$\kappa = (2 \sqrt{2} s_{12} ^c\sqrt{1 -  (s_{12}^c)^2})/(1 - 2 (s_{12}^c)^2)$,
which indeed  leads to $\kappa=4$ for $s_{12}^c=1/\sqrt{3}$.}.
An alternative way of expressing this
{\sl "stationary along a direction" } property  is to
observe that  $\delta s_{23}$ and $s_{13}$ enter
Eqs.(\ref{relative-flavors}) only through one specific combination,
like $\delta s_{123}$ of (\ref{a123-def}). This signals an approximate
degeneracy in the sensitivity to these mixing angles, which would affect
their reconstruction from experimental data.

If second order effects are retained, then dependencies
on all four mixing angle combinations
 $\delta s_{12},  ~\delta s_{23}, ~\delta s_{123} $ and $s_{13}$ appear,
leading to
\bqa
F_e &=&  \frac{1}{3} +
(1 - 2 \sqrt{3} \delta s_{12} + 9 \delta s_{12}^2 - 5 s_{13}^2)
\frac{( 3 F^0_e -1)}{9} \nonumber \\
   & + &
   \Big [(1 - \frac{7}{2} \sqrt{3} \delta s_{12}) \delta s_{123} +
         \delta s_{23} (2 \sqrt{6} \delta s_{12}
         + \frac{4}{9} \delta s_{23} ) \Big ]
(F^0_\tau -F^0_\mu) ~      , \nonumber  \\
\nonumber  \\
F_\mu &=& \frac{1}{2} (1 + \Delta F_{\mu \tau} - F_e)~, \nonumber \\
\nonumber  \\
F_\tau &=& \frac{1}{2} (1 - \Delta F_{\mu \tau} - F_e) ~, \nonumber \\
\nonumber  \\
\Delta F_{\mu \tau} &\equiv& F_\mu - F_\tau = \nonumber \\
     & & \Big [12 \delta s_{123}^2 + \delta s_{23}  ( \frac{68}{9}
     \delta s_{23} -2 \sqrt{ 6} \delta s_{12} ) +
          \delta s_{123} ( \frac{7}{2} \sqrt{ 3} \delta s_{12} -
12 \sqrt{ 2} \delta s_{23} -1 ) \Big ]
( 3 F^0_e -1) \nonumber \\
   & + &
   8 (3 \delta s_{123}^2 - 3 \sqrt{ 2} \delta s_{123} \delta s_{23}
   + 2 \delta s_{23} ^2) (3 F^0_\mu -1)~. \label{relative-flavors-2nd}
\eqa\\

We next turn to the discussion of three
interesting specific cases.

\vspace{0.5cm}
\noindent
\underline{Equipartition condition for fluxes on Earth}.
There is in fact still another reason that makes the specific
point $(s_{23},s_{13})=(1/\sqrt{2}, 0)$ unique in the $s_{23}s_{13}$-plane.
To explain this, we  study in general terms the conditions required
for the initial fluxes and the mixing angles, in order
that  flavor fluxes  on Earth become equal; \ie
$F_e=F_\mu=F_\tau$, a property referred to as flux equipartition.
It is indeed noteworthy that, [besides the Supernovae case
where flux equipartition on Earth is an immediate consequence
of the (approximate) initial flux
equipartition\footnote{This case is discussed in more detail below.}],
the {\sl canonical} case
also leads to equal neutrino fluxes on Earth, provided the neutrino
mixing angles are appropriate.

Using the exact results for large distance oscillations, one finds
 that
\begin{equation}
s_{23}^2=
\frac{1 - F^0_\mu - 2 F^0_\tau}{F^0_\mu - F^0_\tau} +
      \frac{2/3 -  (F^0_\mu + F^0_\tau)}{ (F^0_\tau - F^0_\mu) (1 - s_{13}^2)}
~~, \label{coincidence}
\end{equation}
is  a {\sl necessary} condition\footnote{
Barring some special cases which we have also identified,
where  $(s_{12} = 0, ~ 1/\sqrt{2}, ~1)$ or
$(s_{23} = 0, ~ 1)$ or $( s_{13} = 1)$. Such cases
are clearly excluded by experiment. In particular, the case
$s_{12}=1/\sqrt{2}$ studied in this equipartition context in \cite{Ahluwalia},
has been recently reported  to be   excluded at the  $5\sigma$
level \cite{s12-limits}. }
for equipartition of the neutrino fluxes on Earth.
This condition shows clearly that equipartition requires
a specific correlation between two sets of physically independent
quantities, namely the mixing angles $s_{23}, s_{13}$ and the initial flavor fluxes.
It is noteworthy that since (\ref{coincidence}) is independent
of $s_{12}$, the latter does not have any influence on equipartition,
in agreement with the findings of \cite{Quigg} for  the {\sl canonical case}.

If   $F^0_\mu + F^0_\tau = 2/3$ is now assumed, then Eq.(\ref{coincidence})
implies  $s_{23}^2= (1 - F^0_\mu - 2 F^0_\tau)/(F^0_\mu - F^0_\tau) =1/2$,
which has  been first observed by \cite{Ahluwalia}. If, on top of this,
we impose the  {\sl canonical} initial fluxes
satisfying   Eqs.(\ref{canonical-initial-flavors}), then
$s_{13}  = 0$ is also uniquely  required for equipartition.
Thus, the requirement of exact equipartition of the fluxes on Earth,
combined with  the assumption that the initial fluxes are canonical,
uniquely determines   $s_{23}=1/\sqrt{2}$ and $s_{13}=0$.

The fact  that the initial fluxes
resulting from the astrophysical mechanism, and the experimentally favored
neutrino  mixing angles  {\sl  fall  in the
vicinity of the  unique   equipartition solution} for the fluxes on Earth,
is  a striking coincidence!  This is even more so, if we also remember
the "stationary" property of these fluxes described above. \\

We next turn to the fluxes as determined by the actual ranges of
the mixing angles appearing  in
(\ref{s12-s23-range}, \ref{s13-range}, \ref{a123-range}).
Using the  first order expression (\ref{relative-flavors}),
together with   (\ref{canonical-initial-flavors}), we get
\bqa
&& F_e = \frac{1}{3}(1 ~-~2\,  \delta s_{123})
 ~~ , \nonumber \\
&&  F_\mu=F_\tau  = \frac{1}{3}(1 ~+~\,  \delta s_{123})
~~ ,  \label{canonical-relative-flavors}
\eqa
which agrees with the conclusion of \cite{Yasuda, Equipartition}
that for bimaximal neutrino mixing with very small  $s_{13}$, the
relative neutrino flavor  fluxes are
$F_e \simeq F_\mu\simeq F_\tau\simeq 1/3$.

Our formalism  goes beyond this though, since it also considers
in detail the small deviations from the mixing angle-values
$(\theta_{12} \simeq \pi/5.1,~ \theta_{23} \simeq \pi/4,~ \theta_{13} \simeq 0)$.
To linear order in these deviations, we  find that the
 arriving neutrino fluxes are  independent of
$\delta s_{12}$, and  only depend
on the specific combination of
$\delta s_{23}$ and $s_{13}\cos \delta$ given in  Eq.(\ref{a123-def}).
Moreover,  $F_\mu$ and $F_\tau$  are  always equal and   described
by the "green" line  along the  diagonal of the triangle in Fig.1.

As seen from  Eqs.(\ref{relative-flavors-2nd}) though,
to second  order in
$(\delta s_{12},~ \delta s_{23},~ s_{13})$,
  a weak intrinsic dependence  on
all four parameters $(\delta s_{12},~ \delta s_{23},~ s_{13})$
and $\cos\delta$ appears.
To study this in somewhat more detail, we have reproduced in Fig.1,
the implications of the constraints (\ref{s12-s23-range}, \ref{s13-range}).
The second order formulae in  Eq.(\ref{relative-flavors-2nd})
describes these constraints by  the "blue-plus-red"
 region within the triangle in Fig.1, which almost completely overlaps
  with the "red" region resulted from the numerical analysis
  of the exact expressions   \cite{Yasuda, Equipartition}. As seen there,
 $F_\mu \geq F_\tau$ in the whole allowed  region. This property can
actually be shown analytically, both from the exact formalism or from
the approximation Eq.(\ref{relative-flavors-2nd}) which leads to
\bqa
 F_\mu-F_\tau &=&
   \frac{8}{27} (13 \delta s_{23}^2 + s_{13}^2 \cos^2 \delta +
(\delta s_{23} + s_{13} \cos \delta)^2) ~\geq 0~.
\label{canonical-relative-flavors-2nd}
\eqa

On the basis of this analysis, we find that in the  canonical case
\bqa
&& 0.28 ~ \lsim F_e ~\lsim ~ 0.39 ~~ , \nonumber \\
&& 0.30 ~ \lsim F_\mu ~\lsim ~ 0.36 ~~ , \nonumber \\
&& 0 < \Delta F_{\mu \tau} ~ \lsim 0.073 ~,
\label{Fe-Fnu-Ftau-range-can}
\eqa
and
\bq
0.73 ~ \lsim ~ \frac{F_\mu}{F_e}\simeq (1+3 ~ \delta s_{123})
 ~\lsim  ~1.21 ~~ , \label{Fmu-Fe-ratio-range-can}
\eq
which, as already said,  are consistent with the
results of \cite{Yasuda, Equipartition}.

A virtue of the present derivation, is that  the
effect of a future reduction of the experimental uncertainties on the
mixing angles, can be straightforwardly read analytically
from Eq.(\ref{relative-flavors-2nd}) or even Eq.(\ref{relative-flavors}).
As an example we  note that if  it turns out that \eg
$\delta s_{123}= 0.1$ (compare Eq.(\ref{a123-range})) and that
$F_\mu \simeq F_\tau$,
then the linear formulae  (\ref{canonical-relative-flavors})
should  be  adequate,  leading  to
  $F_e \simeq 0.27$  and $F_\mu \simeq F_\tau \simeq 0.37$. In a future
sufficiently large  neutrino detector such as IceCube \cite{Halzen1},
it might be possible to discriminate this case
from the ideal prediction  $F_e \simeq F_\mu=F_\tau \simeq 1/3 $,
in the TeV-PeV energy range.

\vspace{0.5cm}
\noindent
\underline{Equal initial neutrino fluxes.}
Because of unitarity, if the initial relative flavors
satisfy $F^0_e=F^0_\mu=F^0_\tau$, then the final ones  also obey
$F_e=F_\mu=F_\tau=1/3$, irrespective of the neutrino mixing angles.
This is \eg  the situation roughly  half a second or so after
a supernova collapse and explosion, as suggested by typical
simulations \cite{Raffelt1}.
However, since such configurations can be simulation dependent,
and keeping in mind the
uncertainties related to supernova physics\cite{Raffelt1}, we may parameterize a
deviation in the electron flavor from the initial flux equipartition in the form

\begin{eqnarray}
F_e^0 &=&  \frac{ 1 - 2 \epsilon}{3 } \nonumber \\
F_\mu^0  &=& F_\tau^0 = \frac{ 1 + \epsilon}{3} \label{sn0}
\end{eqnarray}

\noindent
The expected flux differences on Earth are then easily obtained from
Eq.(\ref{relative-flavors}) as

\begin{eqnarray}
F_\tau - F_e &=& (\frac{7}{16} - \frac{3}{2} \delta s_{12} - \delta s_{123}) \, \epsilon
\nonumber
\\
F_\mu - F_e &=& (\frac{7}{16} - \frac{3}{2} \delta s_{12} +  \delta s_{123}) \, \epsilon
\nonumber
\\
F_\mu -F_\tau &=& 2 \delta s_{123}  \, \epsilon   \label{flavdiff}
\end{eqnarray}
\noindent
We call Eqs(\ref{sn0}, \ref{flavdiff}) the "Supernova-type case",
allowing it  to cover also the possibility of TeV   neutrino
sources, which somehow produce  roughly equal neutrino fluxes
for all neutrino and antineutrino flavors.

It is interesting to note that even in the linear approximation in this case,
an initial $\mu/\tau$ flavor equipartition
tends to be removed    by neutrino oscillation effects, as can be seen
from the third equation
in Eq.(\ref{flavdiff}). Moreover,  a simultaneous measurement of the three
flux differences in Eq.(\ref{flavdiff}), provided it is  sufficiently accurate,
 would allow the reconstruction of the three quantities
$\epsilon, \delta s_{123}$ and $\delta s_{12}$.
It should be clear, though, that our analysis is valid only if matter effects
on the observed fluxes can be neglected. Obviously, this would not be the case
for neutrinos which cross the Earth before detection\cite{smirnov}.
Furthermore, star matter effects can be important in
some regions of the supernova \cite{hannestad}.

\vspace{0.5cm}
\noindent
\underline{$F_e^0=1$ case.}
As a last illustration we consider the rather  exotic case where
 $F^0_e=1$, $F^0_\mu=F^0_\tau=0$. In the context of the linear
 approximation formulae Eqs.(\ref{relative-flavors}), we get in this  case
\bqa
F_e &=&\frac{5}{8} \, - \delta s_{12}~~, \nonumber \\
F_\mu &=&\frac{3}{16} +\frac{\delta s_{12}}{2}
\, - \delta s_{123}  ~~, \nonumber \\
F_\tau &=& \frac{3}{16} +\frac{\delta s_{12}}{2}
\, + \delta s_{123}  ~~ , \label{Fe-1-relative-flavors}
\eqa
where, in contrast to the previous situation, the relative neutrino
fluxes have  some sensitivity to $\delta s_{12}$ also. Using
Eqs.(\ref{s12-s23-range}, \ref{a123-range}) we then find
\bqa
&& 0.44 ~ \lsim F_e ~\lsim ~ 0.67 ~~ , \nonumber \\
&& - 0.24 ~ \lsim F_\mu -F_\tau ~\lsim ~ 0.24 ~~ ,
\label{Fe-Fnu-Ftau-range-exotic}
\eqa
where the  uncertainties induced by  $\delta s_{12}$ and $\delta
s_{123}$ are separated.

\vspace{0.5cm}
To summarize, we have studied in this paper some useful analytical
properties of the neutrino flavor fluxes, assuming just three active
neutrino flavors propagating  in the vacuum space
(from the surface of the cosmic sites where they are produced, to Earth)
oscillating among themselves with no neutrino decay
processes\footnote{ The above formulae
can of course be straightforwardly extended to cases including
sterile neutrinos.}.
We  have first emphasized that for any given
$s_{12}$ and $\cos\delta$,
the point $(s_{23}=1/\sqrt{2}~,~s_{13}=0)$
 is {\sl essentially the  unique point}
where (a) all three relative fluxes $(F_e, ~ F_\mu, ~F_\tau)$ on Earth are
equal, when the initial fluxes are canonical, and (b)
a common ($s_{12}$- and $\cos\delta$-dependent) direction exists in
the $s_{23}s_{13}$-plane,
[irrespectively  of the values of the  initial fluxes], along which all three
relative fluxes $(F_e, ~ F_\mu, ~F_\tau)$ on Earth are stationary.
These features have immediate consequences on the sensitivity to the
mixing angles and flavor fluxes. Assuming then that the deviations  of the
neutrino mixing angles from their favored experimental values
$s_{12}=1/\sqrt{3}$, $s_{23}=1/\sqrt{2}$ and $s_{13}=0$ are small, we have
expressed  the observable neutrino fluxes on Earth
in terms of the original ones at the cosmic sites,
keeping  either linear  or quadratic terms
in the above angle-deviations,
and assessed the validity of this approximation through a numerical comparison
with the exact formalism. These
expressions allow to extract information
from Neutrino Astronomy data in a fairly simple way.

\vspace{1cm}
\noindent
\underline{Acknowledgement}\\
One of us (GJG) would like to thank C. Quigg for inspiring
discussions. GM is grateful to G. Sigl for an instructive correspondence.

\begin{figure}[p]
\vspace*{0.cm}
\[
\hspace{-1.cm}\epsfig{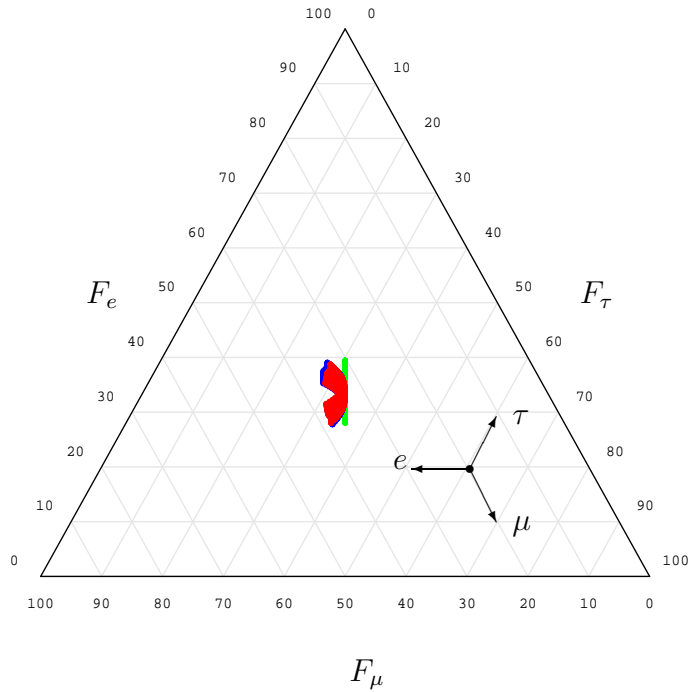}
\]

\vspace*{-5cm}
\hspace*{4cm}{$F_e$}
\hspace*{6cm}{$F_\tau$}

\vspace*{1.8cm}
\hspace*{9.cm}{
\begin{picture}(5, 4)
\put(-1.1, 3){\vector(1, -2){10}}
\put(-1.1, 3){\vector(1, 2){10}}
\put(-1.1, 3){\vector(-1, 0){22}}
\put(-1.1, 3){\circle*{3}}
\put(15, 20){$\tau$}
\put(15, -20){$\mu$}
\put(-30, 3){$e$}
\end{picture}}

\vspace*{2.2cm}
\hspace*{7.5cm}{$F_\mu$}

\caption[1]{ The  relative neutrino flavor fluxes on Earth
$F_e, ~F_\mu, ~F_\tau $ are shown  in the canonical
case $F^0_e=1/3 $, $F^0_\mu=2/3$, $F^0_\tau = 0$. The colored  regions
correspond to mixing angle variations in the intervals
(\ref{s12-s23-range}, \ref{s13-range}) using
{\sl green:} for the linear approximation  (\ref{relative-flavors});
{\sl red:} for the exact formalism result;
{\sl blue plus red:}  for the second order expression
(\ref{relative-flavors-2nd}).
The $F_e$, $F_\tau$, $F_\mu$ coordinates of a specific  point inside the
triangle
are obtained by reading the intersections with the appropriate  axes,
 of  lines emanating  from this specific point  along
 the indicated arrows;  e.g., the illustrated point corresponds to
 $(F_e, F_\mu, F_\tau) = (1/5, 1/5, 3/5)$.}
\label{Diagrams}
\end{figure}

\end{document}